\documentclass[]{interact}
\usepackage{epstopdf}
\usepackage[caption=false]{subfig}
\usepackage{hyperref}
\usepackage{booktabs}

\usepackage{amsmath}

\usepackage[numbers,sort&compress]{natbib}
\bibpunct[, ]{[}{]}{,}{n}{,}{,}

\theoremstyle{plain}

\theoremstyle{definition}

\theoremstyle{remark}

\begin{document}


\title{Improving Biomarker Based HIV Incidence Estimation in the Treatment Era}

\author{
Ian E. Fellows$^1$, Wolfgang Hladik$^2$, Jeffrey W. Eaton$^3$, Andrew C. Voetsch$^2$, Bharat S. Parekh$^2$ and Ray W. Shiraishi$^2$
\newline
\newline
$^1$Fellows Statistics, San Diego, CA, USA \\%
$^2$Division of Global HIV \& TB, U.S. Centers for Disease Control and Prevention, Atlanta, GA, USA \\%
$^3$MRC Centre for Global Infectious Disease Analysis, School of Public Health Imperial College London, London, UK \\%
}

\maketitle

\begin{abstract}
Estimating HIV-1 incidence using biomarker assays in cross-sectional surveys is important for understanding the HIV pandemic. However, the utility of these estimates has been limited by uncertainty about what input parameters to use for False Recency Rate (FRR) and Mean Duration of Recent Infection (MDRI) after applying recent infection testing algorithm (RITA). This article shows how testing and diagnosis in a population reduce both FRR and MDRI compared to a treatment-na\"ive population. Using self-reported testing history, a new method is proposed for calculating appropriate context-specific estimates of FRR and MDRI. The result of this is a new formula for incidence that depends only on reference FRR and MDRI parameters derived in an undiagnosed, treatment-na\"ive, non-elite controller, non-AIDS-progressed population.
\end{abstract}


\subsection*{Information about corresponding author:}
Ian E. Fellows, Ph.D. \\
Fellows Statistics \\
San Diego, California \\
Email: ian@fellstat.com

\subsection*{Funding:}
This research has been supported by the President's Emergency Plan for AIDS Relief (PEPFAR) through the U.S. Centers for Disease Control and Prevention (CDC). Jeffrey W. Eaton was supported by the Bill and Melinda Gates Foundation (OPP1190661), National Institute of Allergy and Infectious Disease of the National Institutes of Health under award number R01AI136664, and the MRC Centre for Global Infectious Disease Analysis (reference MR/R015600/1), jointly funded by the UK Medical Research Council (MRC) and the UK Foreign, Commonwealth and Development Office (FCDO) under the MRC/FCDO Concordat agreement and is also part of the EDCTP2 program supported by the European Union.

\subsection*{Disclaimer:}
The findings and conclusions in this report are those of the authors and do not represent the official position of the U.S. Centers for Disease Control and Prevention.
\\
\\

[Initial draft: 9/14/2021; Completed CDC clearance and peer review: 3/30/2022]

\section{Introduction}

Understanding incidence, the rate at which susceptible individuals in a population become infected, is necessary to effectively monitor and track the spread of an infectious disease and evaluate programs and interventions designed to stop spread.

Cross-sectional surveys combined with assays that can distinguish between recent and long-term infections have been a mainstay for monitoring HIV incidence since early in the pandemic. The basic analytic approach dates back to Brookmeyer and Quinn \cite{brookmeyer1995estimation}, who proposed detection of p24 in the absence of antibodies as a recency indicator (acute infection). Following this, many other indicators were proposed and explored, including less sensitive EIAs \citep{janssen1998new}, CD4 count progression \citep{kaplan1999snapshot}, BED capture EIA \citep{parekh2002quantitative}, avidity index \citep{suligoi2002precision} and most recently, the limiting-antigen (LAg) avidity assay \citep{wei2010development}. 

The analysis framework used to generate incidence from the assay results has remained similar since the method was originally proposed. Kassanjee and colleagues \citep{kassanjee2012new} provided the derivations and formulations most often used today. The formula requires three parameters. The first is a cut-off value for time-since-seroconversion defining the difference between a recent and long-term infection. The second parameter is the rate at which long-term infections test recent on the assay (False Recency Rate, FRR) and the third is the average time a recent individual remains classified as recent by the assay (Mean Duration of Recent Infection, MDRI). Many studies have been performed to estimate these parameters in untreated reference populations \citep{duong2015recalibration, yu2015low, kassanjee2014independent, laeyendecker2018identification,shah2017estimating, parekh2002quantitative, sempa2019performance}. For clarity, in this paper, \textit{reference} FRR and MDRI will refer to the parameters calculated in a reference population. \textit{Residual} FRR and \textit{context-specific} MDRI will be used to refer to the values in the study population after using whatever screening algorithm is employed.

New challenges using recency assays to measure HIV incidence have arisen as the response to the HIV pandemic has progressed to universal ART eligibility for all people with HIV, with widespread adoption of anti-retroviral (ARV) therapy. The recency rate among virally suppressed, ARV-treated and AIDS-progressed individuals is greatly elevated \citep{kassanjee2014independent,shah2017estimating} and the mean duration of a recent infection is also affected by treatment. There is, therefore, a divergence between the parameter values, calculated in untreated reference populations, and the recency rates and durations in the study population.

In order to mitigate the effect of these false recent classifications, The UNAIDS/WHO Working Group on Global HIV/AIDS and STI Surveillance issued a recommendation to treat virally suppressed individuals as non-recent, regardless of their test results \citep{world2011and}. Subsequent guidance suggested that additionally screening out individuals with ARV biomarkers would further avoid elevated recency rates, though this screening has not been formalized into a recommendation \citep{world2015technical}. This approach of combining recency assays with additional biomarker tests for viral load and antiretrovirals has been the standard approach for estimating incidence in Population HIV Impact Assessment surveys \citep{voetsch2021hiv}, the South Africa National HIV Prevalence, Incidence, Behavior and Communication Surveys \citep{woldesenbet2021recent}; and other national household surveys that measure HIV incidence.

The introduction of an additional screening step complicates the analysis of recency assays. While residual FRR is reduced there will still be some misclassification and it is unclear how to adjust the observed proportion classified as recent obtained from a long-term treatment-na\"ive population to get a residual FRR for a combined assay plus screening test. Additionally, because the transition from recent to non-recent classification by the algorithm can either occur as a result of a non-recent result on the assay test or the initiation of ARV treatment, the duration of recency is a competing risk process, which shortens the context-specific MDRI compared to a treatment-na\"ive population.

This paper extends the mathematical framework for incidence estimation from cross-sectional recency assays to the case where individuals are screened out of recency due to treatment initiation. This framing guides us to compute  residual FRR and context-specific MDRI values that are applicable to the population under investigation, accounting for the competing-hazards screening process. Reference FRR and MDRI parameters estimated in treatment-na\"ive, non-elite controller populations are shown to be directly applicable to incidence estimation using a new incidence formula that only requires these as external parameters. Finally, in Section \ref{sec:phia} the new methodology is applied to 11 cross-sectional nationally representative surveys conducted in sub-Saharan Africa from  2015 through 2018.

\section{Recency Equations}

Consider a recency assay that is more likely to classify more recently infected individuals as ``recent'' compared to individuals who have been infected longer. Denote $R$ as a random variable indicating the event where a randomly selected individual in the population is reactive for recency (irrespective of their true timing of infection or treatment status), $T$ (with realization $t$) as the time since HIV seroconversion in years, $I(t)$ as the number of incident cases at time $t$, $H$ as the event that the randomly selected individual is HIV positive, and $q(t)$ as the probability that an individual in the reference population who seroconverted $t$ years ago tests recent on the assay.

If the performance of the recency test does not change over time, and the probability an individual in the study population infected $t$ years ago tests recent is the same as in the reference population ($q(t)$), the rate of infections in the population averaged over $q$ is
$$
\frac{1}{N}\int_{-\infty}^0 I(t)\frac{q(-t)}{\int_{-\infty}^0q(-x)dx}dt = \frac{P(R)}{\Omega},
$$
where $\Omega = \int_{-\infty}^0q(-x)dx$ is the mean duration of that individuals test as recent and $N$ is the population size. Thus the averaged rate of infections is simply the ratio of the proportion of the total population classified as recent over the average duration an individual spends classified as recent.

Adjusting this averaged infection rate for the size of the susceptible (HIV negative) population in order to translate the rate of infections to incidence results in
\begin{equation} \label{eq1}
\bar{\lambda} = \frac{P(R)}{(1-P(H))\Omega}.
\end{equation}
If the susceptible population size is constant, then $\bar{\lambda}$ is the averaged incidence. Otherwise, it can be interpreted as the averaged infection rate adjusted to incidence using the current susceptible population size.

If $q$ is known, Equation \ref{eq1} can be used in conjunction with a cross-sectional survey to estimate $\bar{\lambda}$ by replacing the probabilities with observed sample proportions. Estimation of $q$ among untreated individuals is achieved via a calibration study measuring recency test results in individuals with known time of infection. Many researchers have used such studies to estimate $q$ for various assay types and across geographies (see \cite{kassanjee2017cross} and references therein).

A limitation of these analyses in the context of HIV is that reference data tends to become vary sparse as $t$ gets larger. It is rare, even earlier in the HIV pandemic, for an individual with known infection date to be followed for many years without receiving treatment. Thus, there is considerable uncertainty about the shape of the tail of $q$, and the thickness of the tail can have a large impact on the resulting $\Omega$. Further, if the rate at which long-term infections test recent is constant after some point, $\Omega$ is infinite.

The issue of a potentially infinite $\Omega$ can be solved by truncating $q$ at some cut-off point $\tau$. Hereafter, it is assumed that both the incidence ($\lambda$) and susceptible population size are constant for the last $\tau$ time units. Under these assumptions, $P(T<\tau)$, the proportion of the total population who were infected since $\tau$ is $\lambda\tau(1 - P(H))$, implying that  
\begin{align}
    \lambda &= \frac{P(T<\tau|H)P(H)}{(1-P(H))\tau} \nonumber \\
    P(T \geq \tau|H) &= 1 - \frac{\lambda \tau (1-P(H))}{P(H)}. \nonumber
\end{align}

The probability ($P(R|H)$) that an HIV positive individual is classified as `recent' by a recency assay can be decomposed into those who have been infected for shorter than $\tau$ and longer than $\tau$
$$
P(R|H) = P(R \And T<\tau|H) + P(R \And T\geq\tau|H).
$$ 
Incidence can be expressed as
\begin{align}\label{eq:eq2}
    \lambda &= \frac{P(R \And T<\tau)}{(1-P(H))\Omega_{r}} \nonumber \\
    &= \frac{\big(P(R | H) - P(R \And T \geq \tau | H)\big) P(H)}{(1-P(H))\Omega_{r}} \nonumber \\ 
    &= \frac{\big(P(R | H) - \beta P(T \geq \tau | H)\big) P(H)}{(1-P(H))\Omega_{r}} \nonumber \\
    &= \frac{\big(P(R | H) - \beta\big) P(H)}{(1-P(H))(\Omega_{r} - \beta\tau)},
\end{align}
where $\Omega_{r} = \int_0^\tau q(t)dt$ is the mean recency conditional on $T<\tau$, the reference `mean duration of recent infection' (MDRI), and $\beta=P(R |T \geq \tau, H)$ is the FRR among long-term infections ($T>\tau$). $\beta$ and $\Omega_r$ are not estimable from a single cross-sectional survey, and therefore, appropriate values for FRR and MDRI must be drawn from reference studies.

\section{Adjusting for Recent Infection Testing Algorithms (RITA)}

Given that values must be sourced from reference studies, FRR and MDRI would ideally be stable across populations and time. However, FRR particularly depends on the proportion of the population that is on ARV therapy, who have a high probability of being classified as `recent' even when infected for long durations. Elite controllers have increased false recency rates. In the context of the LAg -Avidity assay, Kassanjee et. al. \cite{kassanjee2014independent} found a 58\% FRR in treated individuals, and a 47\% FRR for those with low viral load. However, they did not find evidence of elevated FRR in those with low CD4 counts. 

In an attempt to correct the classification of these false recents, modern surveys have adopted Recent Infection Testing Algorithms (RITA, \cite{world2011and, world2015technical, voetsch2021hiv}). The most common of these algorithms involve an additional screening component designed to remove treated individuals and elite controllers from those classified as recent infections. Voetsch et. al. \citep{voetsch2021hiv} adopted the nomenclature of RITA2 for a RITA that classifies individuals with ARV biomarkers or viral loads $<1,000$ copies per milliliter. The recency assay may be performed on all screened-in individuals but is more commonly performed on all HIV positive subjects.

This paper utilizes a slightly more restrictive RITA compared to RITA2, which we will call RITA3. In RITA3, individuals are screened out if any of the following conditions hold:
\begin{enumerate}
    \item The subject is HIV negative.
    \item The subject has been previously diagnosed, as determined by either self-report or ARV biomarkers.
    \item The subject is virally suppressed (viral load $<1,000$).
    \item The subject has progressed to AIDS.
\end{enumerate}
The reasons for this more restrictive screening algorithm are threefold. First, in many populations the majority of newly diagnosed individuals immediately initiate ARV treatment and so the difference between screening based on ARVs compared to previous diagnosis may be minimal. Second, false recency rates for cases who were on ARV therapy, but are not currently taking the drugs is unknown and could be potentially higher than never treated cases. Third, it is mathematically convenient to have a process where cases only transition from screened-in to screened-out, and not the other way around. 

If external information on the time between diagnosis and treatment is available, the methods developed here can be extended to the RITA2 algorithm. See Appendix 1 for details.

\begin{figure}
\centerline{\includegraphics[width=1\textwidth]{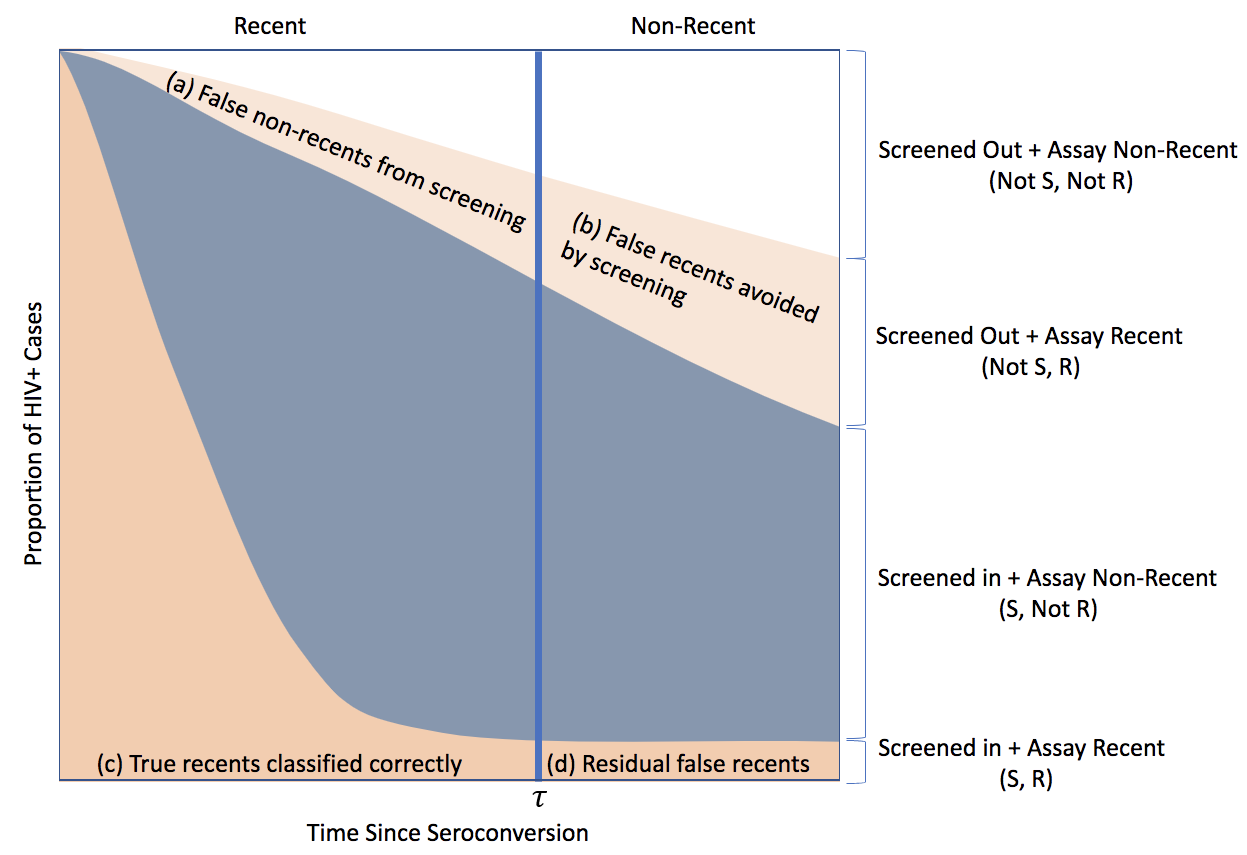}}
\caption{A diagram displaying the interaction between screening and the recency assay and their effect on classification accuracy. \label{fig:dia}}
\end{figure}

Figure \ref{fig:dia} diagrams the decomposed effects of the screening step and assay recency on how individuals are classified by the algorithm. To the left of $\tau$ is the time frame deemed recent, and to the right non-recent. The height of each colored area represents the proportion of cases whose time since seroconversion is $T$ that are screened in/out and/or are assay recent/non-recent. The residual false recency rate is the height of the curve in area (d). If no screening was done, more errors would have occurred, as area (b) would also be included in the residual false recents.

The left side of the plot shows the correctly classified recents in area (c). The area under this curve is the context-specific MDRI. The area (a) represents true recents that are classified as non-recent due to being screened out. Absent screening, the areas of (a) and (c) would be combined and so the effect of the screening process is to reduce context-specific MDRI.


Let $S$ be the event where an individual passes the screening (i.e. they have never been diagnosed, have viral load $\ge 1000$ copies/ml and are, therefore, screened in). Again assuming constant incidence and susceptible population size over the last $\tau$ time units, the probability that an individual infected in the last $\tau$ time units is screened in is
\begin{align*}
    P(S | T<\tau,H) &= \int_0^\tau P(S | T=t,H)P(T=t|T<\tau,H)dt \\
    &= \frac{1}{\tau}\int_0^\tau P(S | T=t,H)dt \\
    &= \frac{\Omega_s}{\tau},
\end{align*}
where $\Omega_s = \int_0^\tau P(S | T=t,H)dt$ is the mean duration of screening in up to $\tau$.

The probability that a screened-in individual is long-term is
\begin{align*}
    P(T \geq \tau | S) &= 1 - P(T<\tau | H,S) \\
    &= 1 - P(S | T<\tau, H) \frac{P(T<\tau|H)}{P(S|H)} \\
    &= 1 - \frac{\Omega_s \lambda (1-P(H))}{P(S \And H)}.
\end{align*}
The above two relationships are useful in the derivations of residual FRR and incidence below.

A standard approach to RITA \citep{voetsch2021hiv} is to simply replace $R$ by $R \And S$ in Equation \ref{eq:eq2}, so that the formula becomes 
\begin{equation}\label{eq:stan}
\lambda =  \frac{\big(P(R \And S | H) - P(R \And S | T \geq \tau,H)\big) P(H)}{(1-P(H))(\Omega_{r,s} - P(R \And S | T \geq \tau,H)\tau)},
\end{equation}
where 
$$
\Omega_{r,s} = \int_0^\tau q(t)P(S | T=t,H)dt
$$ 
is the mean period an individual both tests recent and is screened in up to $\tau$ time units from infection (i.e. context-specific MDRI).
The residual false recency rate ($P(R \And S | T>\tau,H)$) in this equation depends both on how likely an individual is to test recent as well as how likely they are to be screened out. This dual dependence means that it is inappropriate to use the reference FRR as an estimate of this probability. The two effects are decomposed as
\begin{align} \label{eq:betaadj}
    P(R \And S | T \geq\tau,H) &= P(R | T \geq \tau,S,H)P(S | T \geq\tau,H) \nonumber \\
    &=\beta P(T \geq \tau|S,H) \frac{P(S|H)}{P(T \geq \tau |H)} \nonumber \\
    &= \beta \frac{P(S|H)P(H) - \lambda \Omega_s (1-P(H))}{P(H) - \lambda \tau (1-P(H))},
\end{align}
making the assumption that the false recency rate among screened-in individuals $P(R | T \geq \tau,S,H)$ is equal to the reference FRR ($\beta$). It is reasonable to use a reference FRR rate from a population of infections that are known long-term and who also are untreated, have detectable viral loads and have not progressed to AIDS. The fraction to the right of the reference FRR ($\beta$) adjusts for the size of the long-term, screened-in population.

It is now possible to derive a convenient form for the incidence equation, which is identical to plugging Equation \ref{eq:betaadj} into Equation \ref{eq:stan}, as
\begin{align} \label{eq:inc}
    \lambda &= \frac{P(R \And S \And T<\tau)}{(1-P(H))\Omega_{r, s}} \nonumber \\
    &= \frac{\big(P(R | S) - \beta P(T \geq \tau | S)\big) P(S)}{(1-P(H))\Omega_{r,s}} \nonumber \\
    &= \frac{\big(P(R | S) - \beta\big) P(S)}{(1-P(H))(\Omega_{r,s} - \beta \Omega_s)}.
\end{align}
$\Omega_{r,s}$ is a function both of the reference value $q$ and the screening process ($P(S | T=t,H)$), which must be known or estimated from the data.

Equations \ref{eq:betaadj} and \ref{eq:inc} are key relationships that provide insight both into the spread of infection within the population and the dynamics of the RITA algorithm. Equation \ref{eq:inc} provides us with a formulation of incidence that depends only on the reference FRR and not on the more fluid residual FRR. Equation \ref{eq:betaadj}, which itself depends on incidence, defines the relationship between reference and residual FRR as mediated by the screening process.

\section{Estimation}

RITA3 screens out individuals based on their diagnosis status. Estimating $\Omega_{r,s}$ requires the time from seroconversion to diagnosis distribution ($P(S | T=t,H)$). In some populations there may be an external estimate of this. For example, \cite{giguere2021trends} provide time to diagnosis estimates for 40 countries in sub-Saharan Africa. However, it is desirable to be able to obtain it directly from the data.

Time since last HIV test and its result (positive or negative) are collected by most large cross-sectional surveys studying HIV. This information has been leveraged by other incidence methodologies to obtain time from seroconversion to diagnosis distributions \citep{fellows2020new}.

Following \cite{fellows2020new}, it is posited that individuals who engage in regular testing do so at potentially different rates. It is also assumed that individuals who progress to AIDS will become diagnosed due to seeking care for symptoms. For the testing process, it is assumed that the time between HIV tests for each individual $i$ is $f_i(t)$ and that HIV infection is independent of testing. The testing function ($f_i(t)$) is allowed to vary by individual, so high risk individuals may test at a higher rate than low risk individuals. This implies that HIV infection occurs uniformly between last negative and first positive tests. Further, assuming that the timing of the cross-sectional survey is independent of the testing process, the survey also occurs uniformly between the last test, and the next test that the individual would engaged in-if the survey had not occurred. Thus, the empirical distribution of the time since last test among HIV positive individuals with no previous diagnosis is used as an estimate of the time between infection and diagnosis due to regular testing. The estimated survival distribution for time to diagnosis from regular testing as $\hat{d}(t)$.

Time to diagnosis due to AIDS symptoms is modeled as a Weibull distribution with median 10.052 and mean 10.319 years (scale = 1/0.086, shape = 2.516) \citep{brookmeyer1989censoring} and so
$$
A \sim \textrm{Weibull}(\frac{1}{.086},2.516),
$$
where $A$ is the time to AIDS for an individual absent treatment. The survival function is denoted as $a(t) = P(A>t)$.



The probability that an individual has not been screened out at time $t$ is then
$$
\hat{P}(S | T=t,H) = a(t)\hat{d}(t)
$$
and so the mean duration of screening is
$$
\hat{\Omega}_s = \int_0^\tau \hat{P}(S | T=t,H)dt.
$$
The mean duration of recency is
$$
\hat{\Omega}_{r,s} = \int_0^\tau q(t)\hat{P}(S | T=t,H) dt.
$$

If a reliable external measure of time from seroconversion to diagnosis is available, using that would be preferred over the testing history method. Inferring the distribution from testing history requires a number of strong assumptions. One assumption is that individuals transitioning to AIDS become diagnosed. While this may be questionable in practice, the influence this has on $\hat{\Omega}_s$ and $\hat{\Omega}_{r,s}$ is typically minimal. These quantities only depend on the distribution up to time $\tau$, and the AIDS survival curve is relatively flat over the first few years contributing very little to $\hat{P}(S | T=t,H)$.

Constructing $\hat{d}(t)$ assumes that, while an individual's testing behavior may depend on their overall risk-level, it is unrelated to any risk events. In many cases, individuals respond to a potential infection event with testing breaking the uniform distribution of HIV infection between last negative and first positive tests\citep{skar2013towards}. Further, testing history is self-reported and individuals may misreport their last date or misreport whether they have been diagnosed. If misreporting of awareness of HIV status among persons not on ART is of concern, the testing history distribution among HIV negative individuals can be used. This would assume that HIV negative persons test at the same rate as HIV positive persons.

The RITA incidence estimator is obtained by inserting these mean duration estimates into Equation \ref{eq:inc} along with the survey-derived proportion estimates $\hat{P}(R | S)$, $\hat{P}(S)$ and $\hat{P}(H)$
\begin{equation} \label{eq:inc_est}
    \hat{\lambda}_{\textrm{RITA}} = \frac{\big(\hat{P}(R | S) - \beta\big) \hat{P}(S)}{(1-\hat{P}(H))(\hat{\Omega}_{r,s} - \beta \hat{\Omega}_s)}.
\end{equation}
The residual FRR can be similarly estimated by plugging in estimates and proportions into Equation \ref{eq:betaadj}
\begin{equation} \label{eq:resid_frr_est}
    \hat{P}(R \And S | T \geq\tau,H) = \beta \frac{\hat{P}(S|H)\hat{P}(H) - \hat{\lambda}_{\textrm{RITA}} \hat{\Omega}_s (1-\hat{P}(H))}{\hat{P}(H) - \hat{\lambda}_{\textrm{RITA}} \tau (1-\hat{P}(H))}.
\end{equation}
The key advantage of these estimators is that the only external dependencies that they have is for the reference FRR and $q(t)$ estimated in a reference population. The only additional data requirement for the survey is to collect participants' time since last HIV test.

These equations can be generalized in a rather straightforward manner to the case of a less restrictive RITA that only screens out individuals based on ARV biomarkers and viral load. However, additional external information is required on the distribution of times between diagnosis and treatment. See Appendix 1 for details.

\section{The Population-Based HIV Impact Assessment (PHIA)}\label{sec:phia}

The Population-Based HIV Impact Assessment (PHIA) surveys are similarly designed population representative surveys  to estimate HIV prevalence, incidence and viral load suppression in high HIV burden countries in sub-Saharan Africa. In the primary survey analyses, incidence estimation was performed using the LAg-Avidity assay combined with viral load ($>$1000 copies/mL) and absence of ARV in a RITA. Details of the survey methods are presented in \cite{patel2021comprehensive, sachathep2021population}. 

In this section, the PHIA surveys in Cameroon, C\^ote d’Ivoire, Eswatini, Ethiopia, Lesotho, Malawi, Namibia, Rwanda, Tanzania,  Zambia, and Zimbabwe, restricted to respondents aged 15 though 49 years, are reanalyzed. Primary survey analyses assumed that the reference false recent rate was zero and that the reference mean duration was 130 days\cite{voetsch2021hiv}. The Uganda PHIA study was not included in the analysis due to the different reference MDRI used (153 days). Standard errors are calculated using jackknife replicate weights.

The screening process is stricter in the current analysis than in the RITA2 analysis of \cite{voetsch2021hiv} as any individuals with previous diagnosis were screened out, whereas \cite{voetsch2021hiv} only screened out those recently on ARV.

\subsection{$q(t)$ and $\beta$ Parameter Values}

$q(t)$ was constructed by reanalysis of the data from \cite{duong2015recalibration}, who estimated reference MDRI in untreated individuals to be between 130 and 137 days depending on the analysis method used.

\begin{figure}
\centerline{\includegraphics[width=1\textwidth]{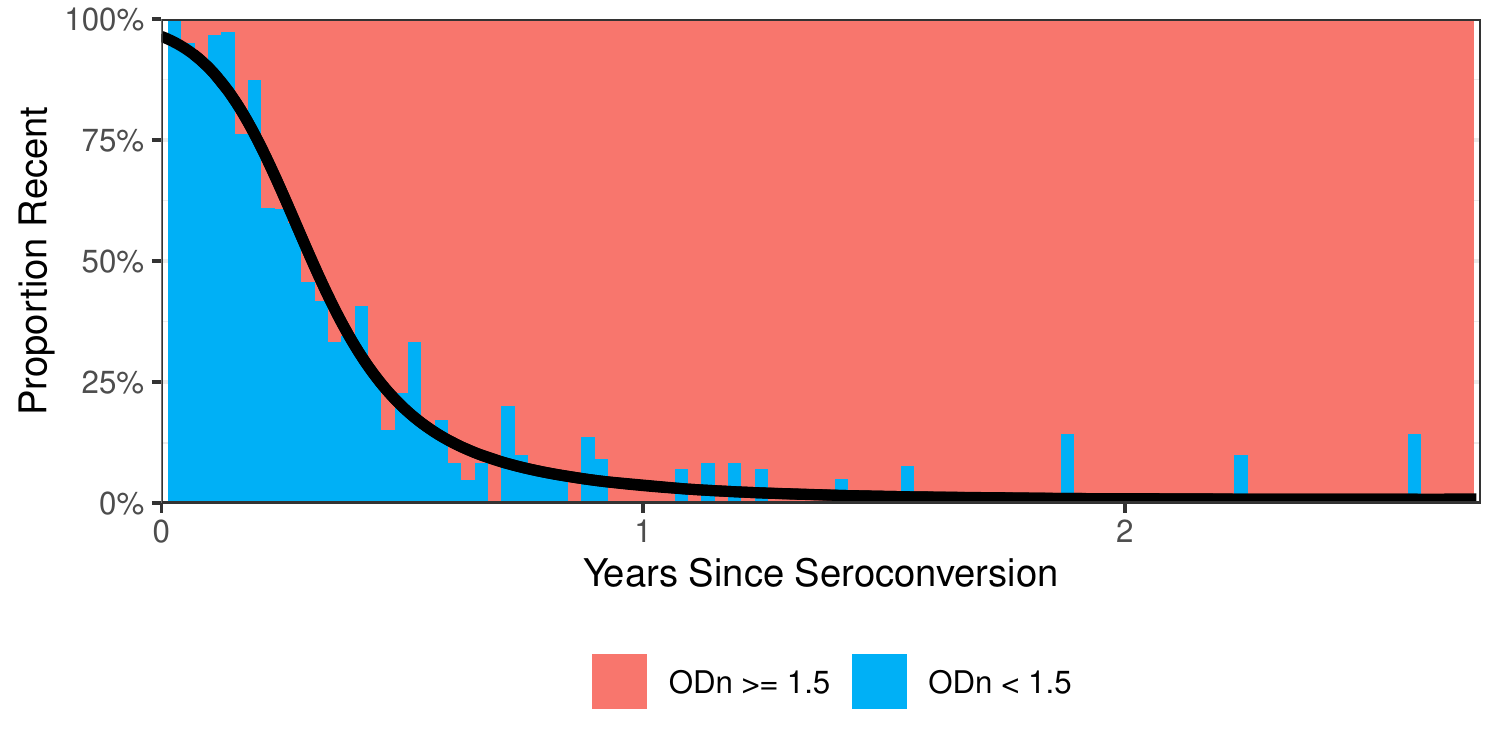}}
\caption{Observed proportion recent by time with fitted $q(t)$ estimated using a generalized additive model (black line).\label{fig:cal}}
\end{figure}

A smoothed generalized additive model (GAM) with a binomial family was used to predict the probability of a recent classification on the assay given time since seroconversion. Figure \ref{fig:cal} shows the observed conditional proportions and the fitted GAM curve.

The estimated MDRI from the GAM curve given $\tau=2$ years is 134.2 days. This is consistent with the seven analyses presented in the originating paper \citep{duong2015recalibration}.

$\beta$ is estimated using the results of \cite{yu2015low}. They found two false recents from among 362 long-term ($>2$ years), untreated individuals with detectable viral loads with CD4 counts $>200$. Based on this, a  reference FRR of $\beta = 2 / 362 = 0.55\%$ was used.

\subsection{Incidence Analysis}

\begin{figure}
\centerline{\includegraphics[width=1\textwidth]{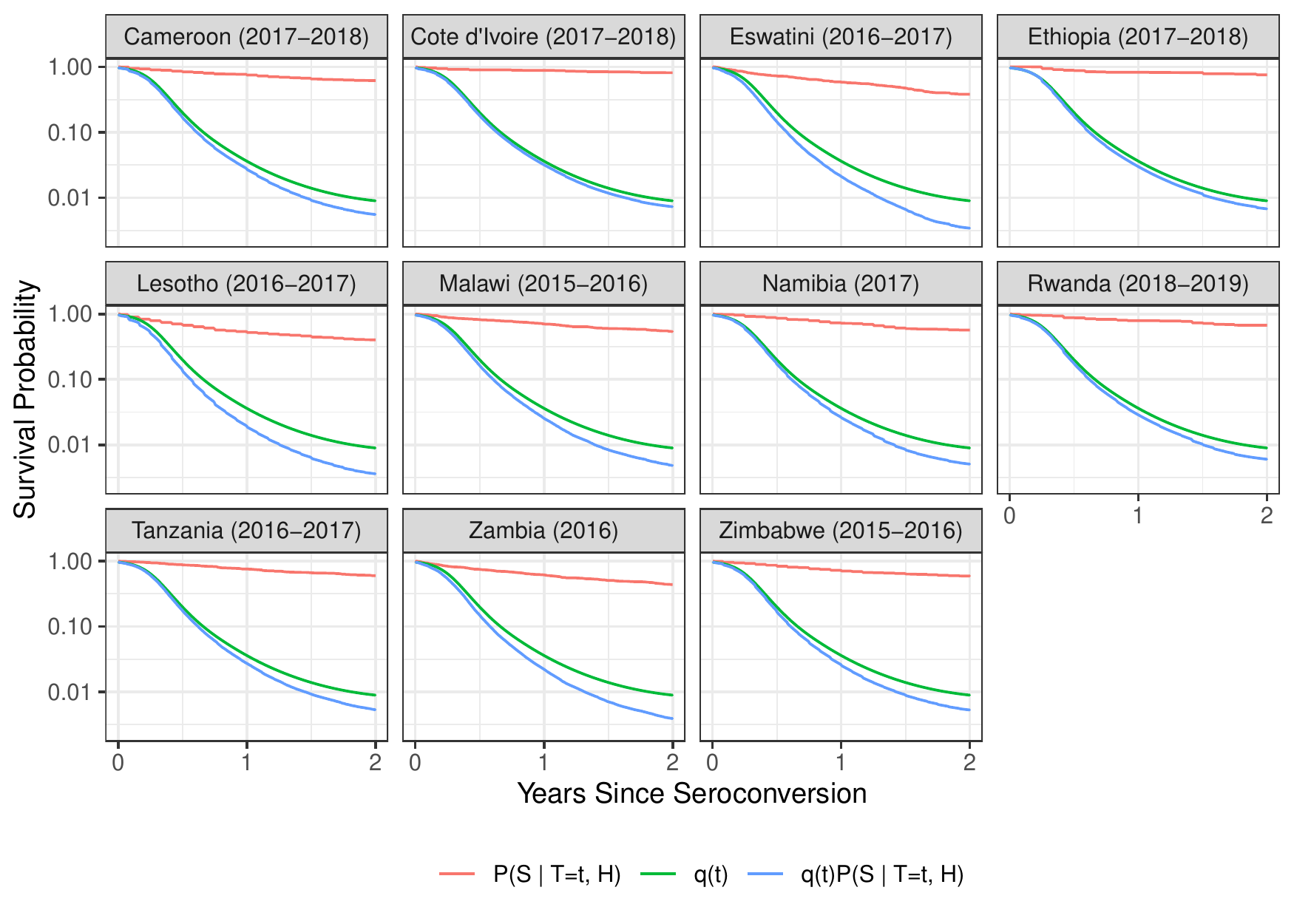}}
\caption{Time to event curves for the competing risk process. Time to diagnosis $P(S|T=t,H)$ is derived from self-reported testing history. Time to LAg-Avidity non-recent classification is defined as the $q(t)$ curve. $q(t)P(S|T=t,H)$ is the probability that an individual is classified as recent by the RITA algorithm given their time since seroconversion. \label{fig:surv}}
\end{figure}

Figure \ref{fig:surv} shows the survival curves in a log scale for each component of the competing risk process. By the end of the recency period (2 years), significant proportions of new positives are predicted to have been diagnosed, with four of the countries reaching greater than 50\% diagnosis rates. 

The probability that an individual is classified as recent is the probability that they are classified as recent on the assay (defined by the $q(t)$ curve) multiplied by the probability that they are screened in (defined by $P(S|T=t,H)$). The result of this is that the survival curve for being classified as RITA recent is below the reference assay curve. Hence, the context-specific MDRI is smaller than the reference MDRI (Table~\ref{tab:par}).

\begin{table}[ht]
\centering
\begin{tabular}{l|rlr|rlr|}
  \hline
  & \multicolumn{3}{c|}{MDRI (days)} & \multicolumn{3}{c|}{FRR (\%)} \\
Country & Reduction & Context-specific (se) & Reference& Reduction & Residual (se) & Reference \\ 
  \hline
Cameroon & -9\% & 122.1 (2.0) & 134.20 & -61\% & 0.21 (0.013) & 0.55 \\ 
  Cote d'Ivoire & -6\% & 125.6 (2.7) & 134.20 & -52\% & 0.26 (0.020) & 0.55 \\ 
  Eswatini & -16\% & 112.3 (2.2) & 134.20 & -91\% & 0.05 (0.005) & 0.55 \\ 
  Ethiopia & -5\% & 127.7 (1.8) & 134.20 & -81\% & 0.10 (0.014) & 0.55 \\ 
  Lesotho & -19\% & 109.4 (1.8) & 134.20 & -85\% & 0.09 (0.005) & 0.55 \\ 
  Malawi & -11\% & 119.8 (1.7) & 134.20 & -83\% & 0.10 (0.007) & 0.55 \\ 
  Namibia & -7\% & 124.2 (1.5) & 134.20 & -89\% & 0.06 (0.008) & 0.55 \\ 
  Rwanda & -7\% & 125.4 (2.4) & 134.20 & -89\% & 0.06 (0.010) & 0.55 \\ 
  Tanzania & -7\% & 125.0 (1.3) & 134.20 & -68\% & 0.18 (0.011) & 0.55 \\ 
  Zambia & -15\% & 113.5 (1.6) & 134.20 & -77\% & 0.13 (0.008) & 0.55 \\ 
  Zimbabwe & -9\% & 122.8 (1.2) & 134.20 & -79\% & 0.11 (0.007) & 0.55 \\ 
   \hline
\end{tabular}
\caption{Recency duration (MDRI) and false recency rates (FRR). Reference value MDRI ($\Omega_r$) and FRR ($\beta$) values are calculated from reference studies in untreated populations. Context-specific MDRI ($\Omega_{r,s}$) and FRR ($\hat{P}(R \And S | T \geq\tau,H)$) values are estimated for the study population after the application of RITA screening. \label{tab:par}}
\end{table} 

The application of RITA reduced the MDRI by between 5\% in Ethiopia to 19\% in Lesotho (Table~\ref{tab:par}). There were large, marked reductions from reference to residual false recency rates. Eswatini had the largest reduction (91\%) and Cote d'Ivoire had the smallest (51\%).

Table \ref{tab:inc} shows incidence estimates under three different values combinations for residual FRR and context-specific MDRI. The ``RITA'' estimate uses Equation \ref{eq:inc_est}, which implicitly uses the adjusted FRR/MDRI values developed in this work. The ``Na\"ive'' estimate assumes that the residual FRR and context-specific MDRI are equal to their reference values. The ``Historical'' estimate sets residual FRR=0 and context-specific MDRI equal to 130; named ``Historical'' because it is the parameter strategy that has been adopted in the past with PHIA surveys \citep{voetsch2021hiv}.

\begin{table}[ht]
\centering
\begin{tabular}{l|rrr|}
  \hline
Country& RITA & Historical & Na\"ive \\ 
  \hline
Cameroon & 0.24 (0.07) & 0.24 (0.07) & 0.19 (0.07) \\ 
  Cote d'Ivoire & 0.02 (0.02) & 0.03 (0.02) & 0.00 (0.02) \\ 
  Eswatini & 1.20 (0.29) & 1.06 (0.23) & 0.51 (0.23) \\ 
  Ethiopia & 0.02 (0.02) & 0.02 (0.02) & -0.02 (0.02) \\ 
  Lesotho & 1.19 (0.26) & 1.04 (0.21) & 0.58 (0.22) \\ 
  Malawi & 0.32 (0.09) & 0.31 (0.08) & 0.15 (0.08) \\ 
  Namibia & 0.35 (0.11) & 0.35 (0.10) & 0.16 (0.10) \\ 
  Rwanda & 0.08 (0.03) & 0.08 (0.03) & 0.04 (0.03) \\ 
  Tanzania & 0.22 (0.05) & 0.23 (0.05) & 0.16 (0.05) \\ 
  Zambia & 0.60 (0.12) & 0.55 (0.10) & 0.37 (0.10) \\ 
  Zimbabwe & 0.37 (0.10) & 0.38 (0.09) & 0.16 (0.09) \\ 
   \hline
\end{tabular}
\caption{Incidence estimates (\%) with different residual FRR and context-specific MDRI values. The ``RITA'' estimate is calculated using Equation \ref{eq:inc_est} and takes into account the estimated reductions in FRR and MDRI from reference to adjusted. The ``Historical'' estimates assume the context-specific MDRI is equal to 130 days and the FRR is 0, a strategy that has been commonly applied in previous publications. The ``Na\"ive'' estimates assume that the residual FRR and context-specific MDRI are equal to their reference values. \label{tab:inc}}
\end{table}

Using the reference values for the residual FRR and context-specific MDRI values (i.e. Na\"ive) results in very low incidence estimates. The point estimate for Ethiopia is negative because the observed proportion recent is lower than the assumed reference FRR value. There is very poor agreement between the RITA estimator and the Na\"ive estimates, with a concordance correlation coefficient of just 61\%. Given that residual FRR is greatly reduced by the RITA screening, using the reference FRR cannot be advised. 

The Historical and RITA estimates are remarkably similar.  The Historical estimate's residual FRR was below that of the RITA estimate, while the MDRI estimate was larger. The combination of these two effects resulted in close agreement and a concordance correlation coefficient of 99\%. A scatter plot of the comparison of the RITA estimates to the Historical and Na\"ive estimates is included in Appendix 2.

\begin{table}[ht]
\centering
\begin{tabular}{l|rrrrrrrr|}
  \hline
  & \multicolumn{8}{c|}{\% Change in $1-\hat{P}(S | T=t, H)$}\\
Country & -100\% & -50\% & -20\% & -10\% & +10\% & +20\% & +50\% & +100\% \\ 
  \hline
Cameroon & -9 & -4 & -2 & -1 & 1 & 2 & 5 & 10 \\ 
  Cote d'Ivoire & -6 & -3 & -1 & -1 & 1 & 1 & 3 & 7 \\ 
  Eswatini & -16 & -8 & -4 & -2 & 2 & 4 & 10 & 23 \\ 
  Ethiopia & -5 & -2 & -1 & -0 & 0 & 1 & 2 & 5 \\ 
  Lesotho & -18 & -10 & -4 & -2 & 2 & 5 & 12 & 28 \\ 
  Malawi & -10 & -5 & -2 & -1 & 1 & 2 & 6 & 13 \\ 
  Namibia & -7 & -4 & -1 & -1 & 1 & 2 & 4 & 8 \\ 
  Rwanda & -6 & -3 & -1 & -1 & 1 & 1 & 3 & 7 \\ 
  Tanzania & -6 & -3 & -1 & -1 & 1 & 1 & 4 & 7 \\ 
  Zambia & -15 & -8 & -3 & -2 & 2 & 4 & 10 & 21 \\ 
  Zimbabwe & -8 & -4 & -2 & -1 & 1 & 2 & 5 & 10 \\ 
   \hline
\end{tabular}
\caption{Percent change in incidence estimate resulting from different percent changes in the probability of being screened out given time since infection ($1-\hat{P}(S | T=t, H)$). \label{tab:sens}}
\end{table}

It is useful to assess how sensitive the incidence results are to the specification of the time to diagnosis distribution as this relies on self-reporting. Table \ref{tab:sens} shows the result of scaling the probability that individuals get screened out ($1-\hat{P}(S | T=t, H)$) on the incidence estimate. For moderate errors within 20\%, there is little effect on the incidence estimates, with the largest error being a 5\% difference. Differences are larger at +/-50\%, but the largest difference (+12\%) is still relatively modest. At +/-100\% misspecification there were significant effects on the incidence estimate, with a maximum change of 28\%, but five of the eleven countries had modest differences of less than $10\%$.

\section{Discussion}

Estimating incidence from cross-sectional surveys has become an important component of our understanding of the HIV pandemic in sub-Saharan Africa. The validity of these estimates has been questioned due to uncertainty around appropriate input parameters for FRR/MDRI. The previous solution, implemented in the PHIA surveys, was to assume there are no false recents (FRR=0) after application of RITA and that MDRI is a constant value across most study areas. However, it is implausible that the RITA screening perfectly removes all false recents and so residual FRR must be some number greater than zero. Additionally, the RITA screening process leads to a context-specific MDRI that is lower than the reference MDRI, and this value varies depending on testing and treatment coverage in the population. Exactly how much larger than zero the residual FRR should be has been an open question, and previous work has not addressed reducing the MDRI to account for the screening process.

This article  developed explicit relationships between reference and residual FRR and context-specific MDRI values, which provided insight into the key drivers that determine how far the adjusted values stray from their reference counterparts due to the screening process involved in RITAs. Using these relations, a convenient formula for incidence was discovered that only depends on external parameters derived in an undiagnosed, treatment-na\"ive, non-elite controller population. The formula is designed to be applied in the context of a cross-sectional survey. Application of these methods to recency assays deployed as part of a routine surveillance system is not recommended at this time but is an area of future research.

The new incidence formula requires an estimate of the time between HIV seroconversion and diagnosis in the study population. This may be obtained from an external source estimate, if it is available. However, following the work of \cite{fellows2020new}, it was demonstrated how self-reported testing history, which is routinely collected in cross-sectional incidence surveys, can be leveraged to estimate time to diagnosis. Thus, no additional data collection or external parameters are required to implement our methodology.

Using testing history to construct the time to diagnosis curve necessarily requires some strong assumptions. Most notably it was assumed that, while high-risk individuals may test more than low-risk individuals, each individual's testing behaviour is independent of infection. Subjects accurately reporting their testing history is also assumed. Fortunately, the sensitivity analysis conducted in the PHIA surveys found that even large 50\% errors in this curve resulted in only modest changes to the resulting incidence estimate. This provides some assurance that the assumptions underlying the utilization of the testing history distribution have limited impact on the resulting estimate.

While this method contrasts with the previous practice of setting residual FRR to zero and setting context-specific MDRI equal to a reference MDRI, the two estimates showed remarkable agreement across the PHIA surveys examined, with a concordance correlation of 99\%. There was some amount of good fortune to this---in the case of these existing PHIA studies, the effect of setting the FRR parameter too low (FRR=0) was counteracted by the effect of setting MDRI too high. The concordance between the two estimates provides additional support for previously published incidence estimates in these PHIA populations.

The agreement between the methods in these datasets is no guarantee of agreement in future studies or of agreement in any sub-populations. Setting FRR to zero when it is known to be greater than zero, as the Historical method does, could potentially lead to statistical criticism of the results. As such, it is recommended to use the methods described here, which explicitly handle the competing risk process underlying the RITA algorithm.

The incidence estimation equation (Equation \ref{eq:inc_est}) is very simple in form; however, there is some computational complexity involved in calculating the context-specific MDRI using the testing history distribution. To facilitate application of the method an R package (``rita'') has been developed, which implements incidence point estimates and uncertainty intervals for the target population, as well as any sub-populations of interest. It is available at https://github.com/fellstat/rita.

\bibliographystyle{plain} 
\bibliography{interacttfqsample}

\end{document}


\section{Appendix 1}

The RITA proposed in this paper (RITA3) classifies all diagnosed and low viral load individuals as non-recent. Previous studies have used RITAs that classify treated individuals as non-recent using viral load and ARV biomarkers (RITA2). This is operationalized as a screening step where individuals are screened out of recency if any of the following conditions hold:
\begin{enumerate}
    \item The subject is HIV negative.
    \item The subject tests positive for ARV biomarkers.
    \item The subject is virally suppressed (viral load $<$ 1,000).
\end{enumerate}

This relaxed RITA fits nicely into our framework. The only difference is that instead of diagnosis and assay recency combining to screen individuals out, treatment and assay recency are combined. Thus, our methods and formulas can be applied to this RITA provided we have an estimate of the probability that an individual infected $t$ time units ago is on treatment. This may be available from an external source, but alternatively we can use the distribution of time from diagnosis to treatment in combination with the distribution of the time from infection to diagnosis. Time from diagnosis to treatment may be available from medical or administrative records, and time from seroconversion to diagnosis can be estimated using the methods in the main paper.

Let $S_{\textrm{Tr}}$ be the event where an individual passes RITA2 screening and $w(t)$ be the probability that an individual is receiving treatment $t$ time units after diagnosis. The probability an individual received their diagnosis at time $x$ and is on treatment at time t is
$$
w(t-x)\frac{d}{dx}(1-\hat{P}(S |T=x,H)) = -w(t-x)\frac{d}{dx}\hat{P}(S |T=x,H)
$$
The estimated probability an HIV+ individual has not been screened out at time $t$ is one minus the probability that they have been diagnosed at some point in the past and are on treatment
$$
\hat{P}(S_{\textrm{Tr}} |T=t,H) = 1+\int^t_0 w(t-x)\frac{d}{dx}\hat{P}(S |T=x,H)dx.
$$
This probability can then be used to calculate mean duration of screening as
$$
\hat{\Omega}^{\textrm{Tr}}_s = \int_0^\tau \hat{P}(S_{\textrm{Tr}} |T=t,H)dt
$$
and mean duration of recency as
$$
\hat{\Omega}^{\textrm{Tr}}_{r,s} = \int_0^\tau q(t)\hat{P}(S_{\textrm{Tr}} |T=t,H) dt.
$$

The incidence equation then becomes
\begin{equation} \label{eq:inc_est}
    \hat{\lambda}_{\textrm{RITA2}} = \frac{\big(\hat{P}(R | S_\textrm{Tr}) - \beta\big) \hat{P}(S_\textrm{Tr})}{(1-\hat{P}(H))(\hat{\Omega}^\textrm{Tr}_{r,s} - \beta \hat{\Omega}^\textrm{Tr}_s)}.
\end{equation}

\section{Appendix 2}

The RITA estimator shows strong agreement with the Historical estimator in the PHIA studies. To better visualize this agreement, it is useful to compare a plot of the RITA estimates with the historical and Naïve estimates. 

\begin{figure}
\centerline{\includegraphics[width=1\textwidth]{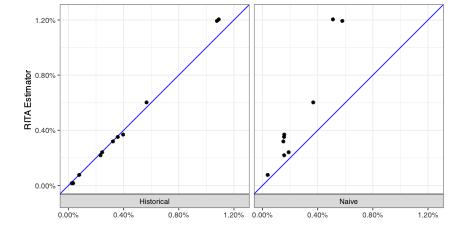}}
\caption{A comparison of the RITA incidence estimates in the PHIA studies to the Naive and Historical estimates. The perfect agreement line is denoted in blue.\label{fig:cal}}
\end{figure}

Figure 1 provides this comparison. Here we see that there is very little deviation from the perfect agreement line (denoted in blue) between the RITA and Historical estimates. The Naïve estimate on the other hand shows large departures, with the larger incidence estimates showing the least agreement.